\documentclass[%
 twocolumn,
superscriptaddress,
 amsmath,amssymb,
 aps,
pra,
longbibliography
]{revtex4-2}

\usepackage{graphicx}
\usepackage{dcolumn}
\usepackage{bm}
\usepackage[T1]{fontenc}

\makeatletter
\newcommand{\overleftrightsmallarrow}{\mathpalette{\overarrowsmall@\leftrightarrowfill@}}
\newcommand{\overrightsmallarrow}{\mathpalette{\overarrowsmall@\rightarrowfill@}}
\newcommand{\overleftsmallarrow}{\mathpalette{\overarrowsmall@\leftarrowfill@}}
\newcommand{\overarrowsmall@}[3]{%
  \vbox{%
    \ialign{%
      ##\crcr
      #1{\smaller@style{#2}}\crcr
      \noalign{\nointerlineskip}%
      $\m@th\hfil#2#3\hfil$\crcr
    }%
  }%
}
\def\smaller@style#1{%
  \ifx#1\displaystyle\scriptstyle\else
    \ifx#1\textstyle\scriptstyle\else
      \scriptscriptstyle
    \fi
  \fi
}
\makeatother

\begin{document}

\preprint{APS/123-QED}

\title{Curie-Weiss behavior and the ``interaction" temperature of magnetic nanoparticle ensembles: local structure strongly affects the magnetic behavior}

\author{Robert E. Camley}
\affiliation{Department of Physics and Energy Science, University of Colorado at Colorado Springs,\\ Colorado Springs CO 80918, United States}
\affiliation{Biofrontiers Center, University of Colorado at Colorado Springs,\\ Colorado Springs CO 80918, United States}

\author{Rair Mac\^edo}
\affiliation{James Watt School of Engineering, Electronics and Nanoscale Engineering Division, University of Glasgow, Glasgow G12 8QQ, United Kingdom}

\author{Karen L. Livesey}
\affiliation{School of Information and Physical Sciences, University of Newcastle, Callaghan NSW 2308, Australia}
\email{Karen.Livesey@newcastle.edu.au}
\affiliation{Biofrontiers Center, University of Colorado at Colorado Springs,\\ Colorado Springs CO 80918, United States}

\date{\today}

\begin{abstract}
In this article, the Curie-Weiss type behavior and the appearance of an ``interaction" or ``ordering" temperature for a collection of magnetic nanoparticles is explored theoretically. We show that some systems where an interaction temperature is reported are too dilute for dipolar interactions to play a role unless at least some of the particles are clumped together. We then show using the most simple type of clumps (particle pairs) that positive and negative interaction temperatures are possible due to dipolar interactions. The clump orientation dramatically changes this result. Finally, we show that an apparent interaction temperature can be measured in magnetic nanoparticle systems that have no interactions between particles, due to some alignment of anisotropy easy axes. These results show that nanoscale physical structures affect the measured magnetic response of nanoparticles.
\end{abstract}

\maketitle


\section{\label{intro }Introduction}

Magnetic nanoparticles are a promising tool for significantly improving human health.  Some uses, such as magnetic resonance imaging (MRI) contrast agents, are already in substantial clinical use.~\cite{rumenapp2012magnetic}  The future involves a number of important applications including targeted drug delivery by using antibiotics attached to magnetic nanoparticles,~\cite{niemirowicz2016magnetic} tumor destruction by heating magnetic nanoparticles through low frequency electromagnetic waves,~\cite{hergt2006magnetic,giustini2010magnetic,deatsch2014heating,balakrishnan2020exploiting} and temperature measurements in MRI in aid of MRI-guided thermal treatments for tumors such as thermal ablation by heating or cooling.~\cite{hankiewicz2019nano} Low-field MRI~\cite{oberdick2023iron} and magnetic particle imaging~\cite{gleich2005tomographic,tay2021superferromagnetic} are emerging imaging technologies that rely on well-characterized magnetic nanoparticles.

Of course, there are issues that must be addressed before widespread in-vivo use.  These include biocompatibility, delivery methods,  and the effect of biological coronas.~\cite{amiri2013protein}     In addition, there are important issues on the physical science side.  Perhaps the most important question is – how do we properly characterize and tailor the physical properties of nanoparticles for particular biological applications?~\cite{maldonado2017magnetic,sandler2019best}  It is exactly that question that we deal with in this paper.

There are multiple parameters that are important for the thermal and electromagnetic behavior of magnetic nanoparticles.  These include their density, the average size of the nanoparticle and the variation in sizes, the average magnetization of a nanoparticle, the average anisotropy of a nanoparticle, etc.  All of these parameters are fundamental in establishing the effectiveness of a given biological application.

One of the key parameters for determining the efficiency of heating tumors by heating magnetic nanoparticles is the magnetic anisotropy, determined by the shape of the nanoparticle, its interactions with other particles, and by its crystalline structure.~\cite{deatsch2014heating,obaidat2015magnetic,gavilan2021size}  Even in ex-vivo experiments, the heating properties of particles may be measured as different from one experiment to another.~\cite{wells2021challenges}  An important and common method of characterizing nanoparticles in solution is a measurement of the ``interaction temperature”, which will be discussed in more detail below.  In this paper we show that the interaction temperature is not generally due to a large-scale interaction, but that a small fraction of particles being in very close proximity to each other may lead to an effect that is measured as an interaction temperature.  Similarly, this measurement may not depend on interactions at all, but could be a consequence of partially aligned anisotropy axes.

We begin with some necessary background in order to show how the interaction temperature arises. Curie's law describes that the low-field susceptibility $\chi$ of a paramagnet scales inversely with temperature $T$.~\cite{coey2010magnetism} The paramagnet is assumed to consist of spins or magnetic moments that are noninteracting. A modification of Curie's law that treats interactions approximately is the Curie-Weiss law,~\cite{mugiraneza2022tutorial} namely
\begin{equation}
    \chi = \frac{C}{ (T-\theta)},
    \label{CurieWeiss}
\end{equation}
where $C$ is a constant and $\theta$ is the so-called ``interaction temperature" or the ``ordering temperature." In the limit of no interactions, $\theta \to 0$ and Curie's law is recovered. 

\begin{figure}[tb]
\includegraphics[width = 0.6\columnwidth]{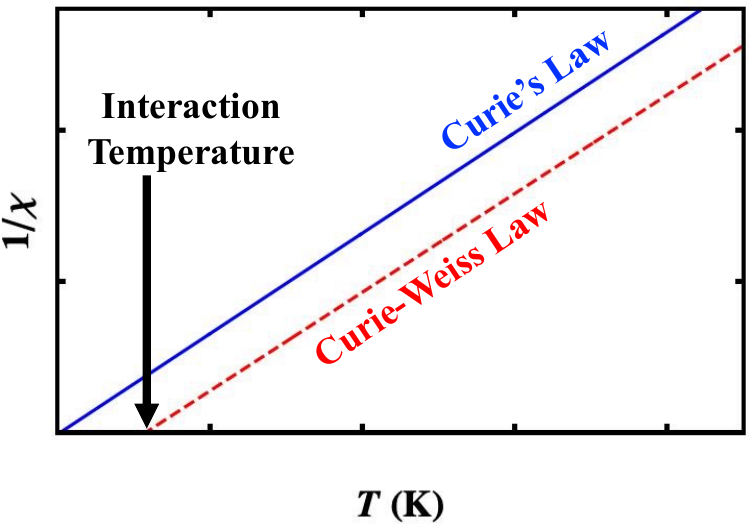}
\caption{ \label{fig:Curie} A schematic showing the inverse susceptibility versus temperature of a paramagnet (solid line, Curies's law). The dashed line shows the Curie-Weiss law (Eq.~\eqref{CurieWeiss}) whereby interactions are included approximately. The interaction temperature or ordering temperature $\theta$ is labelled and corresponds to the temperature axis intercept. Here is is positive indicating ferromagnetic interactions.} 
\end{figure}

Inverting both sides of the Curie-Weiss law, one sees that the inverse susceptibility $1/\chi$ scales linearly with the temperature $T$, and is off-set from the origin by $\theta$. This shape is shown schematically in Fig.~\ref{fig:Curie}. In this way, experimental data for the low-field susceptibility can be used to extract 
$\theta$, which is supposed to give a measure of the strength of the interaction between paramagnetic spins. One should note that Eq.~\eqref{CurieWeiss} is derived using mean field theory, so interactions are treated approximately and their strength is averaged over all contributions (see, for example, Ref.~\cite{coey2010magnetism}). However, the Curie-Weiss law shows strong predictive power for some systems.

This law was used in the 1970s to examine the behavior of spin glasses, where the interactions between spins are frustrated and nontrivial. An example is Ref.~\cite{guy1977gold}, where Fe spins interacts largely by dipolar interactions and a shift in the interaction temperature is recorded as the Fe is made more dilute in the Fe-Au alloy. In fact, the interaction temperature $\theta$ even changes sign, indicating a change in the average interaction from ferromagnetic to antiferromagnetic.

In the 1980s, the superparamagnetic behavior of magnetic nanoparticles was examined through the Curie-Weiss lens.~\cite{soffge1981ac} Instead of atomic spins, one considers the macrospin of an entire single-domain magnetic nanoparticle. Plots of the inverse susceptibility versus temperature again are found to be linear, with an offset that is considered to be due to dipolar interactions between magnetic macrospins. This only occurs in relatively dilute magnetic nanoparticle systems, and for very dense systems the behavior can instead be glassy.~\cite{bedanta2008supermagnetism}

This interpretation of a magnetic nanoparticle ``interaction temperature" is now widely accepted and is discussed in an excellent guide on how magnetic nanoparticles are best characterised.~\cite{maldonado2017magnetic} However, there have been some puzzling experimental results in the literature over the decades which are not yet fully resolved. For example, as early as 1989, Ayoub \emph{et al.} \cite{ayoub1989effect} examined ferrofluids and froze them in different applied field strengths. The idea was to align the macrospin moments and therefore observe a change in the ordering or interaction temperature. Surprisingly, the ordering temperature became more negative (more antiferromagnetic) as the freezing field -- and presumably the degree of alignment -- was increased. The authors supposed that local clusters of particles could give rise to this behavior, but this could not at that time be confirmed. In contrast, in 2012, Urtizberea \emph{et al} performed a similar experiment and saw an increasing, positive interaction temperature with an increase in the alignment field strength.~\cite{urtizberea2012texture}

As another example of a puzzling result, Ref.~\cite{del2009effect} reported the interaction temperature for a series of particle systems where the particles remained the same but their concentration in a polymer matrix was varied. The ferromagnetic interaction temperature at first increased and then decreased as concentration was increased.

Indeed, even the fact that some particle systems record a positive interaction temperature, and others a negative one, is not well understood.

Some attempts have been made to predict the Curie-Weiss behavior of magnetic nanoparticles from nanoscale theories. One of the earliest was a two-dimensional (2D) Monte Carlo simulation showing a positive interaction temperature.~\cite{o1983curie} Some other more recent Monte Carlo studies found a negative interaction temperature.~\cite{chantrell2000calculations,garcia2000influence} However, these works consider nanoparticles located at random, rather than the formation of particle clusters. An article in 2011 notes that ``\emph{to our knowledge, a rigorous, quantitative treatment of magnetic dipolar interaction in the framework of the Curie-Weiss model does not exist at the present time}." \cite{hines2011nuclear}

In this article, we aim to address some of these issues and paradoxical experimental results. We use a variety of theoretical methods to calculate Curie-Weiss behavior for magnetic nanoparticle systems with macrospin moments interacting via dipolar interactions and/or self-anisotropy. We demonstrate that an ``interaction temperature" can occur in situations where particles are anisotropic, without dipolar interactions being present. We also show that many ferrofluid systems are too dilute to display an interaction temperature unless local clusters of particles do in fact form.


Our work indicates that there are multiple  origins for an effective interaction temperature, and makes direct connections with how particle properties and particle configurations affect the net magnetic behavior of samples.  It also indicates the need for   careful experimental measurements of samples that are well characterized at the nanoscale. 

In Sec.~\ref{uniform}, we begin by calculating the interaction temperature for particles at typical concentrations and compare to experimental results. The calculations assume that particles are randomly distributed and show that with these dilute concentrations, no interaction temperature can occur. Next, in Sec.~\ref{pairs}, we consider that particles may form localised clumps. Pairs of particles are considered as a first step to showing how local clumps of particles give rise to an interaction temperature. In these calculations, the anisotropy energy of magnetic nanoparticles is ignored. In Sec.~\ref{aniso}, the effect of anisotropy is considered and dipolar interactions are ignored. Surprisingly, the presence of some net anisotropy in a magnetic nanoparticle system also gives rise to an ``interaction temperature," although particles are not interacting with each other at all. In Sec.~\ref{conc}, conclusions and future work are detailed.




\section{Randomly distributed particles with dipolar interactions}
\label{uniform}

We start by considering a system where $N$ magnetic nanoparticles have random locations in a total volume $\mathcal{V}$. We ignore their single-particle anisotropy energy for now (magnetocrystalline and/or shape anisotropy). Anisotropy will be considered at length in Sec.~\ref{aniso}.  

To find the magnetic susceptibility at a given temperature, an iterative method known as self consistent local mean field (SCLMF) theory is used. It has been used before to find diverse properties such as the thermal magnitude of atomic dipoles in FeRh, \cite{maat2005temperature,mcgrath2020self} thermal skyrmion lattices, \cite{birch2022history} and the thermal behaviour of Fe/Gd multilayers.~\cite{bob87surface} It is not commonly used to model magnetic nanoparticle systems. 

We note that other authors have calculated the thermal magnetization of an ensemble of interacting magnetic nanoparticles using a variety of methods. For example, estimates of the thermodynamic partition function can be made for interacting pairs~\cite{chantrell1983dynamic} and long-range interacting systems.~\cite{landi2013random} This is usually approximate and does not allow one to easily change the configuration of particles from, say, random locations to clumps of particles, as we will explore in this article. Another option is to use Monte Carlo or kinetic Monte Carlo methods.~\cite{ruta2015unified} Finally, one can integrate the stochastic Landau-Lifshitz equation forward in time until equilibrium is reached,~\cite{chalifour2021magnetic} although this is generally computationally demanding for interacting particles.~\cite{haase2012role} We choose SCLMF theory in this Section and the next Sec.~\ref{pairs} as it is computationally cheap and relatively accurate.

The SCLMF method works by assigning initial thermal magnitudes $\langle m_i \rangle$ and directions to the dipole moments $\vec{m}_{i}$, with $i$ ranging from 1 to $N$. The units for the moment $m_{i}$ are Am$^2$. Each particle has a moment of $m = M_s V$ at zero temperature, where $M_s$ is the saturation magnetization and $V$ is the particle volume. Thermal fluctuations mean that $\langle m_{i} \rangle < m$. After initialization, a particle $i$ is chosen at random. Its moment $\langle \vec{m}_i \rangle$ is rotated into the direction of its local effective field $\vec{H}_i$, and then its thermal amplitude is updated using the Langevin function $\mathcal{L}$ according to
\begin{equation}
    \langle m_i \rangle = m ~\mathcal{L}(x) = m \left(\coth(x) - \frac{1}{x} \right),
\end{equation}
with the argument given by
\begin{equation}
    x= \frac{  \vec{m}_{i}  \cdot \vec{H}_{i}}{k_B T }.
\end{equation}
Here $k_B$ is Boltzmann's constant and $T$ is the temperature. 

The effective field of a particle $i$ is given by the external field $H_{\textrm{ext}}$ plus the dipolar fields due to all other particles $j \ne i$ in the system. In other words, 
\begin{equation}
    \vec{H}_{i} = \vec{H}_{\textrm{ext}} + \sum_{j \ne i} \frac{1}{4 \pi} \left( \frac{ 3 \hat{r}_{ij} \left( \langle \vec{m}_j \rangle \cdot \hat{r}_{ij} \right)  }{r_{ij}^3}    - \frac{ \langle \vec{m}_j \rangle }{r_{ij}^{3} }  \right) ,
\end{equation}
where $r_{ij}$ is the center-to-center spacing between particles $i$ and $j$, with $\hat{r}_{ij}$ a unit vector pointing from particle $j$ to particle $i$. For simplicity, the external field is taken to be in the $z$ direction.

This process of picking a particle, and updating the direction and magnitude of $\langle \vec{m}_{i} \rangle$ is repeated until there is no change in the system. At this point, the net volume magnetization $M_{\textrm{tot}} = \left(\sum_{i} \langle m_{i}^{z} \rangle \right)/(NV) $ -- units of A/m -- along the applied field direction is recorded for a given value of the external field $H_{\textrm{ext}}$ and temperature $T$. The volume here $NV$ is the total volume of magnetic material. Note that some research groups use mass magnetization, in which case the units are different but where the inverse susceptibility intercepts the temperature axis (see Fig.~\ref{fig:Curie}) should not change.

The susceptibility can then be calculated from the net volume magnetization. Small magnetic fields are typically chosen (less than 10~Oe) so that  $M_{\textrm{tot}}$ is linear with field. Also, the temperature must be sufficiently large so that the thermal energy swamps the magnetic energy meaning that $M_{\textrm{tot}} \to 0$ when $H_{\textrm{ext}} \to 0$ (i.e. there is zero remanence).~\cite{maldonado2017magnetic} If both these conditions are met, then the susceptibility is given by
\begin{equation}
    \chi (T) = \frac{ M_{\textrm{tot}} }{ H_{\textrm{ext}} },
\end{equation}
which is unitless in SI units. For lower temperatures, where the dipolar interactions become important compared to thermal fluctuations, then the magnetic nanoparticles may fall into a ``glassy" or frustrated local energy minimum.~\cite{bedanta2008supermagnetism} A low-field susceptibility still exists but since there is a magnetic remanence, it should be calculated according to:
\begin{equation}
    \chi (T) = \frac{ M_{\textrm{tot}}^{(1)} - M_{\textrm{tot}}^{(2)} }{ H_{\textrm{ext}}^{(1)} - H_{\textrm{ext}}^{(2)} },
    \label{betterChi}
\end{equation}
where the superscripts (1) and (2) represent two values of the external applied field, which are both weak. Although the glassy state is not the main focus of this article, signatures of it appear in our results.

With the method explained, we begin by examining a well-characterised system from the literature, assuming that the magnetic nanoparticles are located at random positions.

As an example, we take parameters for 7~nm-radius magnetite particles from Ref.~\cite{maldonado2017magnetic}. To match the particle density $c_0 =0.000054$\% by volume (corresponding to 0.1\% Fe$_3$O$_4$ in PDVD polymer by mass), we place $N=312$ particles at random in a 2~micron cubed simulation space (ensuring they do not overlap in space). Each particle is modelled as a single magnetic dipole located at its center with a maximum magnitude $m = M_s V$, with $M_s = 430$~kA/m the saturation magnetization and $V=4 \pi r^3 /3$ the particle volume. The thermal magnetic moment of each particle $\langle \vec{m}_{i} \rangle$ is calculated using the SCLMF method detailed in this section. It is confirmed that the net magnetization $M_{\textrm{tot}}$ scales with the strength of the applied magnetic field for the temperatures used ($T >10$~K) with anisotropy ignored so these particles are certainly in the superparamagnetic regime.

\begin{figure}[tb]
\includegraphics[width = 1.0\columnwidth]{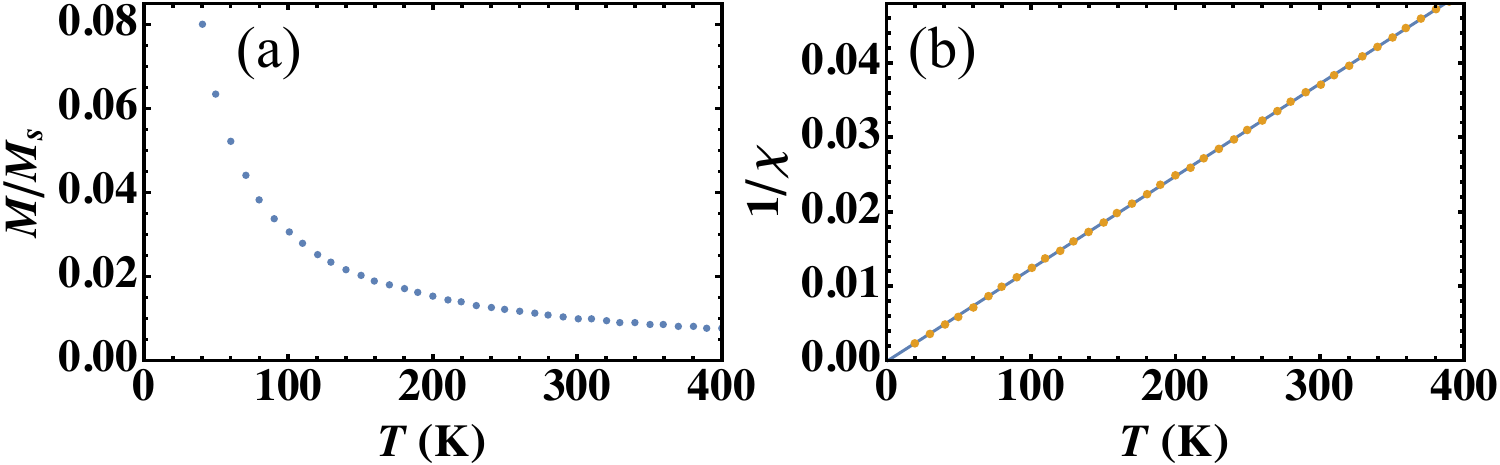}
\caption{ \label{fig:random} (a) Normalized magnetization and (b) inverse susceptibility (unitless) versus temperature, calculated using SCLMF method for a dilute, random system of 7~nm-radius magnetite nanoparticles. Anisotropy is ignored for now and the particles are interacting. The mass concentration of magnetite is 0.1\% (0.0054\% by volume) and $M_s = 430$~kA/m, following Ref.~\cite{maldonado2017magnetic}. An interaction temperature is absent in (b), where the line shows a best-fit to the data between 200 and 400~K.} 
\end{figure}

It is found that the inverse susceptibility versus temperature is linear, with an intercept given by $\theta =0.01$~K, found by fitting to data between 200 and 400~K. The (a) magnetization and (b) inverse susceptibility data are plotted versus temperature in Fig.~\ref{fig:random} in the interest of completeness, although the interaction temperature calculated here is essentially zero. This shows that if the magnetic nanoparticles in this system are randomly placed then they are far too dilute for dipolar interactions to give rise to an interaction temperature. However, in Ref.~\cite{maldonado2017magnetic} these particles are reported to have an interaction temperature $\theta = 42$~K. We provide two mechanisms for which this interaction temperature may arise. In Sec.~\ref{pairs} we show that a small amount of local clumping of particles in such dilute systems can give rise to an interaction temperature which is sensitive to the nanoscale-configuration of the cluster. In Sec.~\ref{aniso} we show how magnetic anisotropy may result in an apparent interaction temperature.

\begin{figure}[tb]
\includegraphics[width = 0.8\columnwidth]{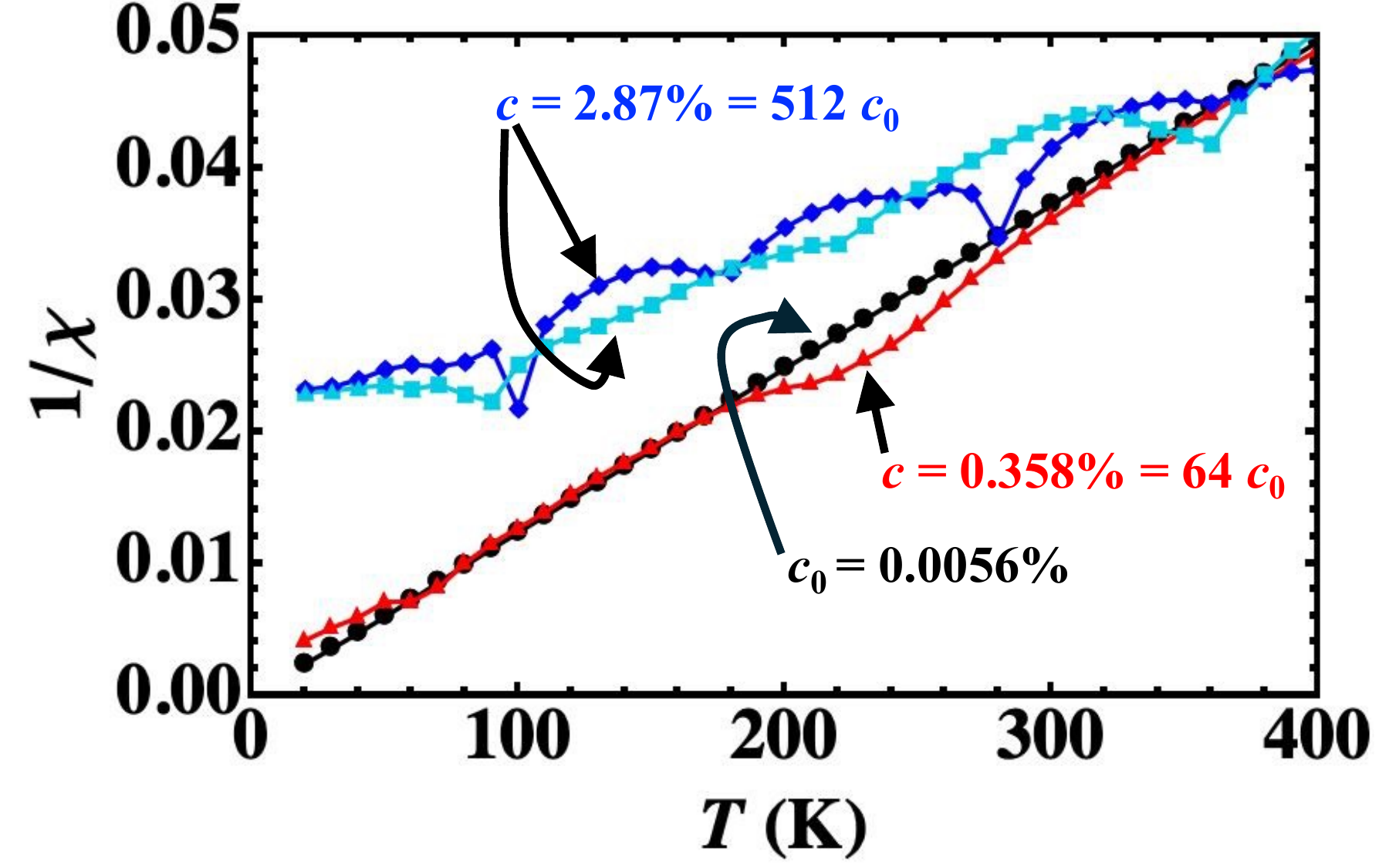}
\caption{ \label{fig:denser} Inverse susceptibility versus temperature for three different volume concentrations of magnetic nanoparticles: $c_0 = 0.0056$\% (black circles), $c= 64 c_0$ (red triangles) and $c=512 c_0$ (blue squares and diamonds). There are two curves for the densest concentration, to show that the random positioning of particles changes the magnetic response.} 
\end{figure}

Before moving to these two sections, we explore making the random system of particle less dilute (or more dense). This is similar to experimental studies where the concentration of magnetic nanoparticles is varied,~\cite{del2009effect} although in most experiments the nanoscale configuration (and presence or absence of clusters) is not known. Some results for different particle concentrations are given in Fig.~\ref{fig:denser}. Again, inverse susceptibility is plotted versus temperature. The black data (straight line) is for randomly placed particles at the original, dilute, volume concentration $c_0 = 0.0058$\%. The red triangles are for a concentration $c=0.358$\%, which is 64 times denser. This is achieved by still using $N=312$ particles, but shrinking the cubic simulation volume by 64 times so as it is 0.5~$\mu$m on each side. One sees that the inverse susceptibility still looks to intercept the temperature axis near the origin, but is no longer a straight line. This is due to the dipolar interactions, and is discussed below for an even denser concentration.

In Fig.~\ref{fig:denser} there are two example curves corresponding to a very dense system with volume concentration of magnetic material $c=2.87$\%, which is 512 times larger than the most dilute system. These are denoted by the blue squares and triangles, respectively. Depending on the randomly chosen initial configuration of particles, the results vary, as is seen by the difference between these two curves. 
However, with the increased particle density, there is generally a reduction in the slope of the inverse susceptibility curve. In addition, there is a general increase in the magnitude the inverse susceptibility. Essentially what happens is that the dipole interaction becomes so strong so that the dipole moments are basically locked into place and the system can not respond to the external magnetic field. As a result, the susceptibility is decreased and the inverse susceptibility becomes larger and less dependent on temperature.
For both the $c=2.87$\% cases, there is a large, negative interaction temperature. Putting a linear fit through the data (from 400~K to 0~K) results in $\theta = -273$~K and -226~K, respectively. 

Notice that both these curves display some dips in the inverse susceptibility, on top of the overall linear trend. These are due to frustrated dipolar interactions in the system. Occasionally, as the system is iterated upon using the SCLMFT method and as the temperature is varied, a single particle will reverse it's moment, leading to a somewhat sharp change in the magnetic response. Changes in the random positioning of particles therefore alter where the dips occur.

Note that we have iterated 10,000$\times N$ times using SCLMFT to achieve the results shown in Fig.~\ref{fig:denser}. We also tried 20,000$\times N$ iterations and the results change a little, but the random starting configuration is by far more important in determining the magnetic response. 

There is clearly a transition somewhere between a volume concentration of 0.358\% and 2.87\% which leads to both a negative interaction temperature and also a nonlinear dependence of the inverse susceptibility on temperature. For the densest system, it is also important to calculate the susceptibility as given in Eq.~\eqref{betterChi}.

Ref.~\cite{bedanta2008supermagnetism} gives a review of the transition from superparamagnetic behavior for very dilute particle systems to glassy and finally ``superferromagnetic" behavior for very dense systems.

\section{Local interaction temperature: pair examples}
\label{pairs}

We have seen in the previous section that global dipolar fields are unlikely to create an ``interaction temperature” signature in the plot of inverse susceptibility as a function of temperature because the dipolar fields at extended distances are simply too weak.  However, in this section we illustrate that the dipolar fields from nearby nanoparticles can cause such a signature, even if the percentage of particles that are paired is relatively small.

\begin{figure}[tb]
\includegraphics[width = 0.8\columnwidth]{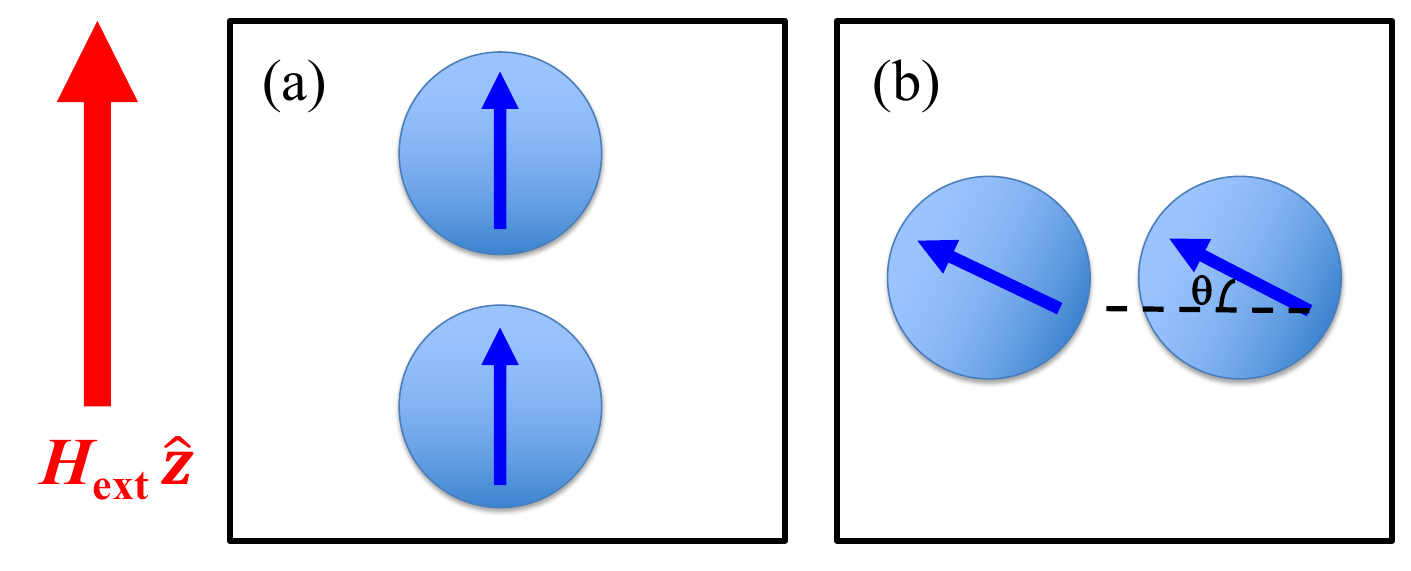}
\caption{ \label{fig:pairs} An illustration of two possible configurations for pairs of nanoparticles. (a) Particles positioned along the field direction (``vertical pair"), and (b) particles positioned perpendicular to to the external field $\vec{H}_{\textrm{ext}} = H_{\textrm{ext}} \hat{z}$ (``horizontal pair"). The arrows inside the spherical particles indicate the likely orientations for the particle macrospin moments in a small external field. For the horizontal pair, the moments are at a small angle $\theta$ from the horizontal, which is approximately equal for each particle.} 
\end{figure}

To illustrate this, we consider two cases illustrated in Fig.~\ref{fig:pairs}:
\begin{itemize}
\item a pair of nanoparticles oriented along the external field direction (``verticle pair");
\item a pair of nanoparticles oriented perpendicular to the external field direction (``horizontal pair")
 \end{itemize}
We place the particles close enough, with their surfaces a few nanometers from each other, so that the local dipolar fields are substantial. 

In this section we only consider dipolar interactions between particles where the anisotropy can be neglected. (Again, the effect of anisotropy is considered in Sec.~\ref{aniso}.)  In this case one may treat the system with the SCLMF theory detailed in the last section.  

We consider two spherical magnetite nanoparticles, each with a radius of 7~nm, where the center to center distance is 20~nm.  The saturation magnetization is a little higher than in the previous section, with $M_s =4.5\times 10^5$~A/m.  At this distance the effective dipolar field of one particle acting on the other at zero temperature is about  $\mu_0 H = 0.015$~T which is quite large compared to the typical external field of 0.0001~T (10~Oe). However, at $T = 400$~K, the external field and the local dipolar field are about the same magnitude.  This, temperature-dependent difference in the field magnitudes leads to the ``interaction temperature” signature.  

\begin{figure}[tb]
\includegraphics[width = 0.8\columnwidth]{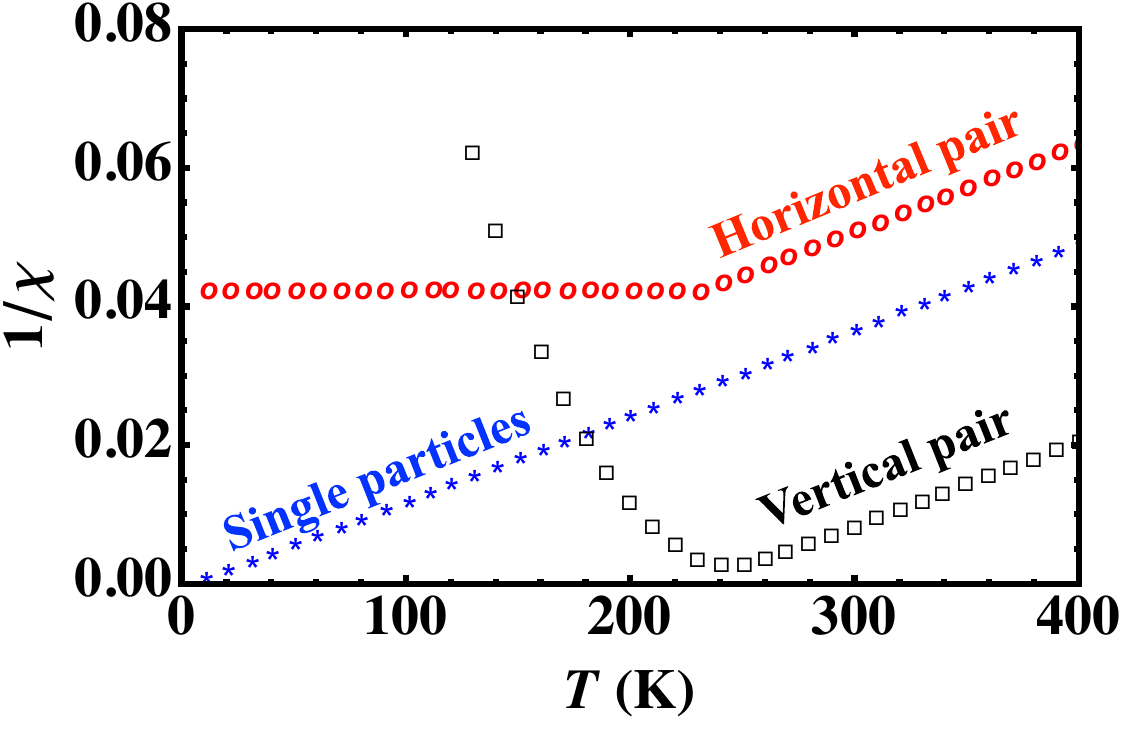}
\caption{ \label{fig:pairInv} Inverse susceptibility versus temperature, calculated for 7~nm-radius single particles ($*$ marker) and for horizontal pairs of particles (circles) and vertical pairs (squares). The parameters used are based on magnetite and are given in the main text.} 
\end{figure}

We calculate results for the inverse susceptibility for a single particle, and for the pairs of particles oriented both horizontally and vertically.  
The results of these calculations are shown in Fig.~\ref{fig:pairInv}. At temperatures above around 250~K, all three calculations show Curie-Weiss behavior with $1/\chi$ being linear.

The single-particle case ($*$ marker) shows the typical straight line intersecting the origin that is characteristic of an isolated particle with a temperature dependence given by the Langevin function.  As the temperature is reduced, the thermal average moment on the particle increases, increasing the susceptibility and decreasing the inverse susceptibility.  

The pair results in Fig.~\ref{fig:pairInv} are more interesting with a distinct change in inverse susceptibility around 220--230~K.  We can understand one feature of these results immediately.  The vertical pairs (squares) have a much lower inverse susceptibility than the horizontal pairs (circles) at high temperatures.  This is a result of the primarily parallel orientation for the two magnetic moments in the vertical pair.  Their dipolar field enhances the net magnetization and the susceptibility, leading to a smaller inverse susceptibility. Fitting the data between 300--400~K with a line and extrapolating back to the axis reveals an interaction temperature of +230~K for these vertical pairs, which is large compared to most experimental measurements of real systems.

The bend in the inverse susceptibility for the pairs is less easy to understand.  First, we discuss the vertical pair situation.  At high temperatures (400~K) the induced dipole field is about the same magnitude as the applied field and increases as the applied field is increased. As the temperature is decreased, the dipole field becomes stronger because it is created by the thermal averaged magnetic moment of one of the particles.  This enhances the magnetization of the other particle increasing the susceptibility and reducing the inverse susceptibility.  Finally, around 220~K the dipole field is about 20 times larger than the external field. At this point, increasing the external field creates almost no change in the net magnetization, leading to a small susceptibility and a large inverse susceptibility.

The horizontal pair behavior is more complex.  At high temperatures the magnetic moments of the particles line up with the external field.  This means the dipolar field of one particle acting on the nearby particle is opposite to the external field, reducing the net field felt by the second particle.  This results in a smaller effective field acting on the single-particles and, as a result, the magnetization change with field is small leading to a small susceptibility and a larger inverse susceptibility. (See Fig.~\ref{fig:pairInv})  Around 230~K, the thermally averaged magnitude of the magnetic moments has grown substantially larger,  and it is now energetically favorable for the system to transition to a configuration close to that seen in Fig.~\ref{fig:pairs}(b), where the magnetization is primarily horizontal.  Surprisingly, the inverse susceptibility is nearly constant in this region as a function of temperature. We present a simple calculation outlining the source of this behavior in the Appendix.

The SCLMFT calculation method can calculate inverse susceptibility for all temperatures, but it is only at high temperatures that the Curie-Weiss law is valid and a linear fit is put through the data to estimate the interaction temperature. For the horizontal pairs, a fit through the 300--400~K data yields an interaction temperature of $-111$~K. This indicates antiferromagnetic interaction, which is indeed the case and was discussed above. Note that the vertical and horizontal pairs have an interaction temperature differing by 340~K. This shows that it is not just the formation of a clump of particles that is important for determining the interaction temperature, but also that clump's orientation with respect to the external applied field which is important.

Real systems, of course, are likely to be composed of single particles as well as pairs or other collections of particles.  We explored the consequences of this for different filling fractions 
of the three primary categories (single, horizontal pairs, vertical pairs). It was found that the horizontal pairs of particles have a much weaker effect on the interaction temperature than the vertical pairs. 

\begin{figure}[t]
\includegraphics[width = 0.8\columnwidth]{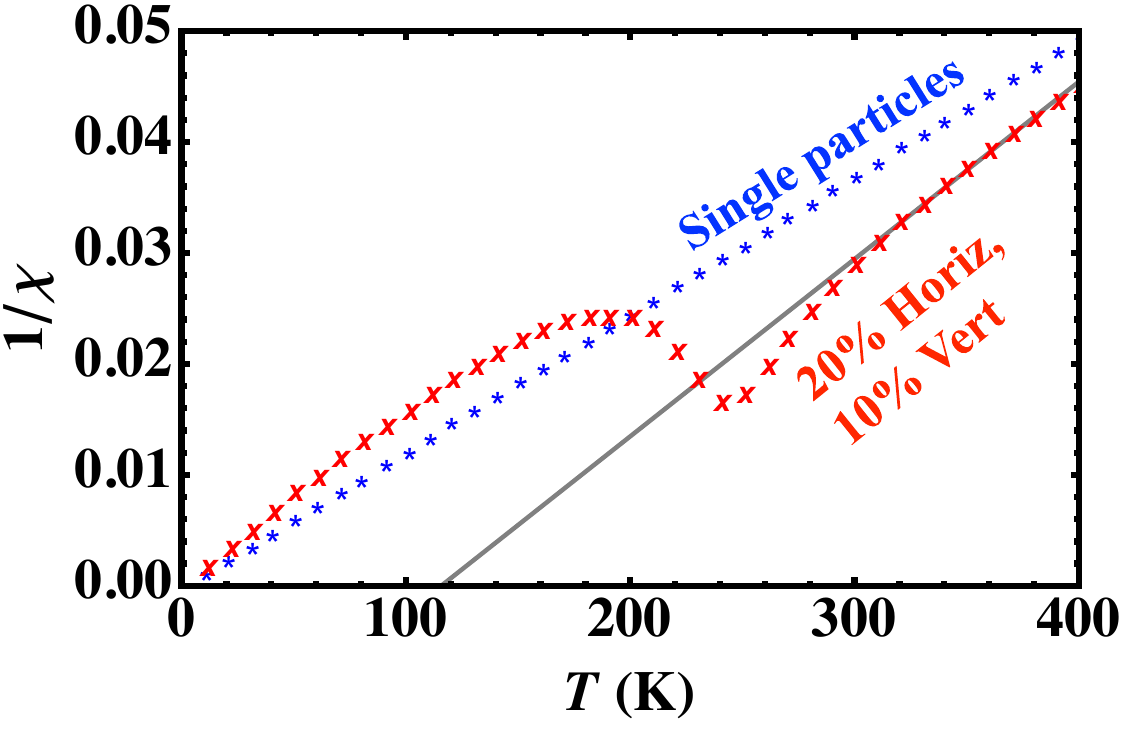}
\caption{ \label{fig:pairInvMix} Inverse susceptibility (unitless) versus temperature for an ensemble of 7~nm-raidus magnetic nanoparticles where 70\% are isolated (single), 20\% are in horizontal pairs, and 10\% are in vertical pairs (crosses). The result for single particles only is shown again as a guid ($*$ markers). A linear fit (gray line) is made to the data between 300--400~K and extrapolated back to the temperature axis.} 
\end{figure}

Although many different filling fractions were examined, in Fig.~\ref{fig:pairInvMix} one instance is shown where 70\% of particles are single, 20\% are in horizontal pairs and 10\% are in vertical pairs. 
One sees that fitting the high temperature data (300--400~K) gives an interaction temperature of +116~K. Although there are more horizontal pairs than vertical pairs, the interaction temperature remains positive. This is a signature of the fact that the dipolar interactions are stronger for the vertical pairs (Fig.~\ref{fig:pairs}(a)) than for the horizontal pairs (Fig.~\ref{fig:pairs}(b)) at high temperatures. Also, although only 10\% of particles are in vertical pairs, the interaction temperature of the ensemble is measured to be a large value of +116~K.

A recent study \cite{mamiya2021evaluation} showed that the interaction temperature for a series of 7.8~nm diameter magnetite nanoparticles with different silica shells was under 10~K. However, first order reversal curve (FORC) analysis showed that there were strong antiferromagnetic and ferromagnetic interactions between the magnetic cores when closely packed together. This suggests that the antiferromagnetic and ferromagnetic interactions may in some cases almost cancel each other out. This is consistent with what is shown here for simple particle pairs. Imagining how our results would carry over to larger clumps where individual particles have both vertical and horizontal neighbours and next-nearest neighbours, this cancellation of interaction temperature seems plausible.


On this point, it is worth mentioning that the strength of interparticle dipolar interactions can also be estimated using other experimental techniques. While the Weiss interaction temperature $\theta$ involves a static magnetic field, many dynamic measurements -- such as FORC \cite{mamiya2021evaluation} --reveal more information about local energy barriers. In particular, Shtrikman and Wohlfarth used an interaction temperature $T_0$ to alter the N\'eel relaxation time of magnetic nanoparticles that are weakly interacting, using a Vogel-Fulcher law.~\cite{shtrikman1981theory} Other models also exist for how interactions change the magnetic relaxation time of particles,~\cite{dormann1988dynamic} and the superparamagnetic Blocking temperature.~\cite{morup1994superparamagnetic} While all these dynamical ways to probe dipolar interactions are beyond the scope of this article, we note that often the Weiss interaction temperature measured using inverse susceptibility measurements is assumed to be the \emph{same} temperature used to estimate the reduction in N\'eel relaxation times (i.e. $T_0 = \theta$), \cite{maldonado2017magnetic} which in turn affects estimates of magnetic anisotropy energy barriers.~\cite{livesey2018beyond}

In summary, we have seen in this Section that simply introducing a small number of paired nanoparticles can produce an interaction temperature $\theta$ in the inverse susceptibility.  Within this model, the measured interaction temperature indicates more about the distribution and local configuration of particles rather than any properties of the particles themselves, or any long-ranged global interactions.

\section{Anisotropy induced ``interaction" temperature}
\label{aniso}

While in the last subsection anisotropy energy was ignored and we explored the effect of dipolar interaction energy between particles, in this subsection the reverse is true. We show that an ``interaction" temperature can occur in a system of noninteracting magnetic nanoparticles that have a distribution of anisotropy easy axis directions. We follow a method similar to that given in Refs.~\cite{raikher1983magnetization,chantrell1985low}, whereby the thermodynamic partition function is used. Raihkher~\cite{raikher1983magnetization} calculated the magnetization for a system with anisotropy texture in 1983. 

Note that the SCLMF method used in the previous sections becomes less reliable when there is anisotropy because a particle's energy landscape is reduced to just two possible magnetic moment directions corresponding to the two energy minima. The SCLMF method can therefore be appropriate in the very large anisotropy energy barrier limit. Because of this, we found it more appropriate to use a partition function method which shows agreement with a Monte Carlo simulation that we also tested.  We note, however, that the SCLMF produces results similar to the general trends obtained in this section, i.e. that an interaction temperature can be caused by anisotropy alone in some circumstances. 

\begin{figure}[tb]
\includegraphics[width = 0.7\columnwidth]{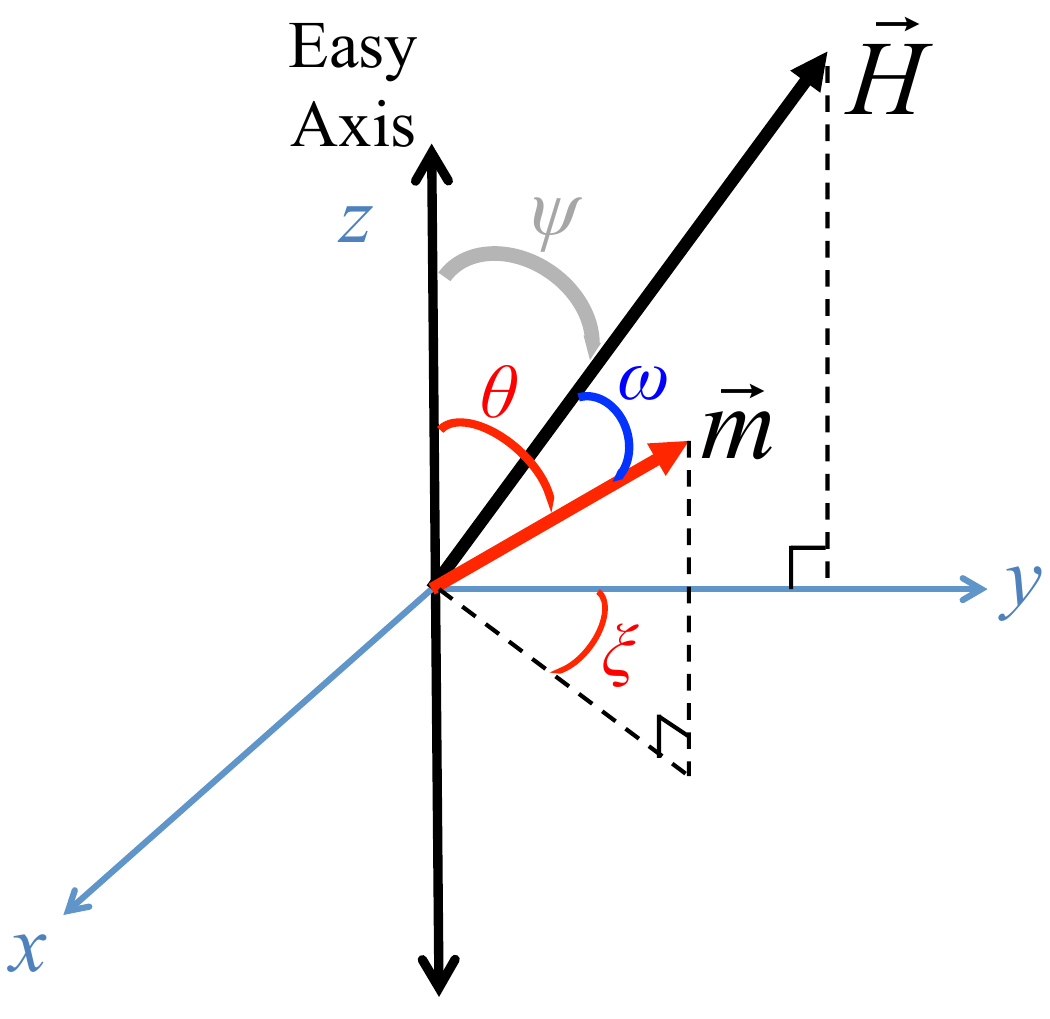}
\caption{ \label{fig:ang} The geometry of all angles considered in the calculation of the magnetic partition function. The easy axis is vertical. The applied field $\vec{H}$ forms an angle of $\psi$ with the easy axis, while the magnetization $\vec{m}$ of a magnetic nanoparticle is at an angle $\theta$ from the easy axis. The angle between $\vec{H}$ and $\vec{m}$ is $\omega$, which can be calculated in terms of $\psi$, $\theta$ and azimuthal angle $\xi$. Note that an azimuthal angle is typically defined in spherical polar coordinates from the $x$-axis, so this definition is different but consistent with that used in Ref.~\cite{chantrell1985low}.} 
\end{figure}

Magnetic nanoparticles are considered with uniaxial anisotropy characterised by energy density $K$. A weak external magnetic field $H$ is applied in order to induce a magnetic response, and measure a susceptibility.  Each particle may have a different angle $\psi$ between its easy axis and the applied field direction. The energy of a particle with it's moment $\vec{m}$ forming an angle $\theta$ with the easy axis direction is given by
\begin{eqnarray}
    E(\psi, \theta, \xi) &=& -KV \cos^2 \theta  \\
    && - \mu_0 M_s V H \left[ \cos\theta \cos\psi +\sin\theta \sin\psi \cos\xi \right], \nonumber
    \label{energy}
\end{eqnarray}
where $V$ is the particle volume, $\mu_0$ is the permeability of free space, and $\xi$ is the azimuthal angle between the applied field and the moment. The geometry is shown in Fig.~\ref{fig:ang} for a single instance of $\vec{H}$ and $\vec{m}$.

The partition function is found by summing up all possible angles for the magnetic moment in a particle and is given by
\begin{equation}
    Z = \int_{0}^{\pi} d\theta \int_{0}^{2 \pi} d \xi ~\sin \theta~ e^{-E(\psi,\theta,\xi)/(k_B T)},
\end{equation}
where $T$ is the temperature and $k_B$ is Boltzmann's constant. The thermal, dimensionless magnetic moment of a particle with a given angle $\psi$ between the field and the easy axis is then
\begin{equation}
    \langle m  (T,\psi) \rangle = \frac{1}{Z} \int_{0}^{\pi} d\theta \int_{0}^{2 \pi} d \zeta~ \sin \theta\cos\theta e^{-E(\psi,\theta,\zeta)/(k_B T)},
    \label{thermalm}
\end{equation}
which depends on the temperature. The angled brackets here denote a thermal average. Without the anisotropy energy contribution to Eq.~\eqref{energy}, Eq.~\eqref{thermalm} gives the well-known analytic Langevin function given in Eq.~\eqref{lang}. 

Much work has been done to create analytic series expansions for the partition function of magnetic nanoparticle systems in various limits.~\cite{garcia2000statics} However, here we will solve the integrals numerically.

To model an ensemble of magnetic nanoparticles where the easy axes lie at various angles with respect to the applied field direction, one can sum Eq.~\eqref{thermalm} over all possible angles $\psi$. Let $w(\psi)$ represent the weighting function or the probability density for the angle $\psi$ between the easy axis and the field. Then, the thermal magnetic moment of the ensemble is 
\begin{equation}
    \langle m _{\textrm{ens}} (T) \rangle =  \int_{0}^{\pi} d\psi ~w(\psi) \sin \theta ~\langle m  (T,\psi) \rangle,
    \label{thermalmEnsemble}
\end{equation}
where Eq.~\eqref{thermalm} must be substituted in. The result is a triple integral, but software packages such as \emph{Mathematica} can handle these with relative ease.

The most commonly assumed distribution of easy axes is that they are random in the ensemble. In other words, the probability density $w(\psi) = 1$ (uniform) for all angles. In this case, one finds that Eq.~\eqref{thermalmEnsemble} recovers the result of the regular Langevin function for weak fields, namely:
\begin{equation}
    \langle m _{\textrm{unif.}} (T) \rangle =  \coth(x) - \frac{1}{x},
    \label{lang}
\end{equation}
where $x=\mu_0 M_s V H/(k_B T)$. In other words, the presence of anisotropy energy barriers in each magnetic nanoparticle has no effect on the net magnetization of the ensemble when the easy axes point in random directions, as recognized by others.~\cite{shliomis1993frequency,garcia2000statics} The unitless susceptibility in this low-field, high-temperature limit is then simply
\begin{equation}
    \chi = (M_s/H) \times \langle m _{\textrm{unif.}} (T) \rangle.
\end{equation}

However, one may consider a ``textured" system of easy axes (as called in Ref.~\cite{chantrell1985low}) where the weight $w(\psi)$ is not uniform. This is a realistic scenario to consider, especially if a magnetic nanoparticle sample is prepared in a field, or in a system with some mechanical strain, such that the easy axes have some overall alignment. Here, we will eventually consider a system with some weak alignment, as may be typical in real systems. But first we consider the extreme scenario where all the easy axes are \emph{aligned} with the magnetic field ($\psi=0$, or $w(\psi) = \delta(\psi)$). We will call this the ``aligned case" and show that an ``interaction temperature" is seen in plots of inverse susceptibility $1/\chi$ versus temperature $T$.

We consider $M_s = 430$~kA/m, $K=13,600$~J/m$^3$, $H=10$~Oe ($\mu_0 H = 1$~mT) and particle radius $r=3$~nm. The external field is very weak ( $\mu_0 M_s V/(k_B T) = 0.009$ at $T=$400~K) to ensure the susceptibility is linear with field. For these parameters, the blocking temperature is calculated to be $T_B \sim KV/(25 k_B ) = 4.5$~K.~\cite{livesey2018beyond} This means that our analysis of superparamagnetic behavior should be at temperatures well above 5.7~K, and we choose 300~K and higher. (Note that the partition function method used here can be applicable at lower temperatures near the Blocking temperature,~\cite{garcia2000statics} but then $1/\chi$ does not scale linearly with $T$.)

\begin{figure}[tb]
\includegraphics[width = 0.8\columnwidth]{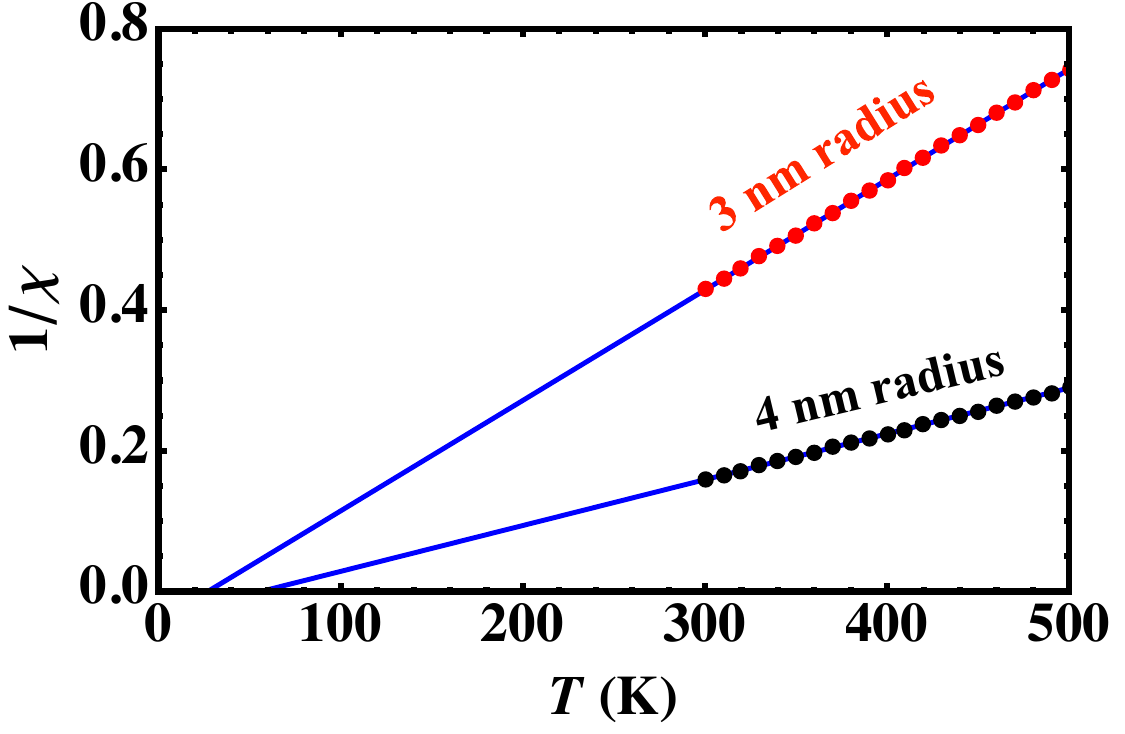}
\caption{ \label{fig:align} The inverse susceptibility versus temperature for 3~nm-radius particles (red dots) and 4~nm-radius particles (black dots), with easy axes aligned with the probing external magnetic field. All other parameters are given in the main text. The dots are calculations based on Eq.~\eqref{thermalm} with $\psi=0$. The blue lines are a linear fit through the data. The ``interaction temperature" is $+27$~K (3~nm radius) and $+56$~K (4~nm radius).} 
\end{figure}

Fig.~\ref{fig:align} shows a plot of inverse susceptibility $1/\chi$ as a function of temperature $T$ for the aligned case, with $r=3$~nm (red dots) and $r=4$~nm (black dots). The susceptibility is found using Eq.~\eqref{thermalm} for the dimensionless moment with $\psi=0$, and using $\chi = (M_s / H) \times \langle m  (T,0)\rangle $. Linear fits through the data (blue lines) extrapolate back to $\theta=+27$~K (3~nm radius) and $+56$~K (4~nm radius). This is significant as the ``interaction temperature" occurs in a system where there are no inter-particle interactions. Moreover, the larger particles give rise to a larger interaction temperature, presumably as they have a larger anisotropy barrier.

Perfect alignment of easy anisotropy axes in a nanoparticle ensemble is impossible to create in a thermal system. However, the easy axes can be partially aligned, for example, by preparing the sample in a biasing external field. This was indeed attempted in Ref.~\cite{ayoub1989effect,urtizberea2012texture} (as mentioned in the introduction) with the interaction temperature becoming more negative or positive as the degree of alignment is increased. Obviously, this is in contrast to the calculation shown here, where a positive interaction temperature was found for maximum alignment. As pointed out in Ref.~\cite{ayoub1989effect}, their negative interaction temperature may be due to local particle interactions. In an experiment on particles in a fluid, the applied field used to align the easy axes may also aid in pulling particles close together and so the effects of both anisotropy texture and local dipolar interactions are inextricably linked. On the other hand, in simulations one can deconvolve the two effects, as we have done here.

We also looked at particles with partial alignment of the easy axes. To do this, we chose a random number from a uniform distribution $y \in [-0.5,0.5]$, then calculated $\psi = \arccos(0.5) - \arccos(y)$ randomly for 5000 particles. This creates a distribution of easy axes that is peaked in the applied field direction and extends down to a maximum of 60$^{\circ}$ from the pole. The integral in Eq.~\eqref{thermalm} is then calculated numerically for the 5000 instances and averaged to get the ensemble average. For this distribution and $r=3$~nm, the interaction temperature was +21~K, reduced from +27~K in the fully aligned case quoted above. 

Sitbon \emph{et al.}~recently explored how magnetic anisotropy causes deviations from Langevin-type behavior when nanoparticles are in a large field and immobilized in a solid, even if their easy axes are distributed randomly.~\cite{sitbon2020effects} Systems with no preferred easy axis direction were also found to deviate from Langevin-type behaviour at temperatures just above the Blocking temperature, since then the anisotropy energy barrier is still large in comparison to thermal fluctuations.~\cite{wiekhorst2003anisotropic} Our work here is in a different regime to these two cases, with temperatures very high compared the Blocking temperature and applied fields very weak. However, all these studies show that the magnetocrystalline anisotropy needs to be considered when one studies magnetic nanoparticles in the superparamagnetic regime.



\section{Conclusions}
\label{conc}

In this work we have presented calculations for magnetic nanoparticle systems so as to reveal some issues associated with the so-called ``interaction temperature." First, we have showed that systems with an experimentally-measured interaction temperature are too dilute for the interaction temperature to arise if the particles are positioned at random. On the other hand, an interaction temperature can occur if the particles are somehow clumped together. By looking at the most trivial clumps (pairs of particles) we show that the interaction temperature can be either positive or negative depending on the clump's shape and orientation with respect to the externally applied field.

That first calculation of interacting particle pairs in Sec.~\ref{pairs} ignores the effect of magnetocrystalline anisotropy. In Sec.~\ref{aniso}, we instead remove interparticle interactions and consider the effect of magnetocrystalline anisotropy on the interaction temperature. We show that an interaction temperature can occur in ``textured" ensembles of magnetic nanoparticles. That is, in systems where the uniaxial easy axes have some degree of alignment and do not point in random directions. For the example of 3~nm-raidus magnetite particles shown here, the interaction temperature has a maximum value of +27~K for full alignment of easy axes along the external field direction.

This work shows that a detailed understanding of nanometer-scale structures in magnetic nanoparticle systems is required in order to correctly interpret the meaning of the static interaction temperature. Magneto-crystalline anisotropy, dipolar interactions between particles in a clump, and even longer-ranged dipolar interactions in relatively dense particle systems can all contribute to the interaction temperature and may indeed partially counter each other out. This may be why there is a wide range of reports on both positive and negative interaction temperature values between relatively similar systems.

The shape anisotropy of particles is presumably also very important and would be interesting to include as future work. In addition, it would be interesting to calculate the interaction temperature for realistic, large clumps of particles (rather than the particle pairs used here). Particle clumps could be assumed from, say, TEM images. They could also be predicted via simulations of particles moving in fluid, though the calculated formation of complicated clumps of particles that match realistic systems is complicated by size polydispersity~\cite{ivanov2007magnetic} and the effects of ligand friction.~\cite{anderson2021simulating} Moreover, one would want to know not only the physical location of particles in formed agglomerates, but also the directions of the particles' easy axes in order to correctly predict the Curie-Weiss offset temperature $\theta$.

\begin{acknowledgments}
The work of R.E.C. was supported by the Australian-American Fulbright Commission through a Distinguished Chair position at the University of Newcastle, Australia. The School of Information and Physical Sciences at the University of Newcastle is acknowledged for their travel support. For the purpose of open access, the authors have applied a Creative Commons Attribution (CC BY) licence to any Author Accepted Manuscript version arising from this submission.
\end{acknowledgments}

\appendix

\section{Analytic estimate for horizontal pairs}
\label{analytic}
It was noted in Sec.~\ref{pairs} that the inverse susceptibility is constant at low temperatures for the horizontal pairs of particles (see Fig.~\ref{fig:pairInv}, red circles). In this Appendix we give a brief calculation showing why this is the case.  

As pointed out in the text, at low temperature, the magnetization of the particles is nearly horizontal because of the large dipolar field (see cartoon in Fig.~\ref{fig:pairs}(b) which shows a particle pair).  We can write the energy of a particle as
\begin{equation}
    E= - \vec{m} \cdot \vec{H} = - \left( m_z B_z + m_x B_x \right).
    \label{EApp1}
\end{equation}
Here $\vec{m}$ is the net thermal magnetic moment of a particle, $B_z$ is the external field used to create the susceptibility (labelled $H_{ext}$ in Fig.~\ref{fig:pairs}), and $B_x$ is the nearly horizontal dipolar field created by one particle acting on the other. The geometry is illustrated in Fig.~\ref{fig:pairs}(b), and $\theta$ is the angle of the net magnetic moment with the horizontal axis. 
 
Because the particle magnetizations are nearly horizontal, we can use the small-angle approximation, that is
\begin{equation}
    m_x \sim m \left( 1 - \frac{ \theta^2 }{2} \right),~~~~~~~m_z \sim m \theta.
\end{equation}
Now, suppose $B_x$ scales with the moment magnitude $m$, i.e. as the temperature decreases, $m$ and $B_x$ change by the same factor.   This is expected since the horizontal field,  $B_x$, is produced by the magnetic moment of the neighboring particle. Then we have $B_x=\alpha m$ but  $B_z$ doesn’t change.
The energy Eq.~\eqref{EApp1} becomes
\begin{equation}
    E = - B_z m \theta - m^2 \alpha \left( 1 - \frac{ \theta^2 }{2} \right).
\end{equation}

The minimum energy is found by setting $\frac{\partial E}{ \partial \theta} =0$.  Then solving for  $\theta$ one obtains
\begin{equation}
    \theta = \frac{ B_z }{ m \alpha }.
\end{equation}
The net moment in the direction of the applied field is then given by
\begin{equation}
    m_z = m \theta = \frac{ B_z }{ \alpha },
\end{equation}
and the susceptibility is 
\begin{equation}
    \chi = \frac{ d m_z }{ d H_z } = \frac{ \mu_0}{\alpha} = \textrm{constant} .
\end{equation}
So, the susceptibility and inverse susceptibility are now both constant at low temperatures. This was seen in the SCLMFT results presented in Fig.~\ref{fig:pairInv}.

\bibliography{apssamp}

\end{document}